# Injection Locking and Coupling Dynamics in Superconducting Nanowire-based Cryogenic Oscillators

Md Mazharul Islam[1], *Graduate Student Member, IEEE*, Md Shafayat Hossain[2], Kathleen E Hamilton[3], and Ahmedullah Aziz[1*], *Senior Member, IEEE*

[1]Dept. of Electrical Eng. and Computer Sci., University of Tennessee, Knoxville, TN, 37996, USA
[2]Dept. of Materials Science and Engineering, University of California, Los Angeles, CA 90095-1595, USA
[3]Oak Ridge National Laboratory, Oak Ridge, TN 37831, USA
*Corresponding Author. Email: aziz@utk.edu

*Abstract*— Oscillators designed to function at cryogenic temperatures play a critical role in superconducting electronics and quantum computing by providing stable, low-noise signals with minimal energy loss. Here we present a comprehensive numerical study of injection locking and mutual coupling dynamics in superconducting nanowire (ScNW)-based cryogenic oscillators. Using the design space of standalone ScNW-based oscillator, we investigate two critical mechanisms that govern frequency synchronization and signal coordination in cryogenic computing architectures: (1) an injection locking induced by an external AC signal with a frequency near the oscillator's natural frequency, and (2) the mutual coupling dynamics between two ScNW oscillators under varying coupling strengths. We identify key design parameters—such as shunt resistance, nanowire inductance, and coupling strength—that govern the locking range. Additionally, we examine how the amplitude of the injected signal affects the amplitude of the locked oscillation, offering valuable insights for power-aware oscillator synchronization. Furthermore, we analyze mutual synchronization between coupled ScNW oscillators using capacitive and resistive coupling elements. Our results reveal that the phase difference between oscillators can be precisely controlled by tuning the coupling strength, enabling programmable phase-encoded information processing. These findings could enable building ScNW-based oscillatory neural networks, synchronized cryogenic logic blocks, and on-chip cryogenic resonator arrays.

## I. INTRODUCTION

The growing demand for scalable quantum computing, high-sensitivity deep-space instrumentation, and energy-efficient high-performance computing has accelerated the development of cryogenic electronics—particularly superconducting electronics [1]–[4]. Operating at ultra-low temperatures, superconducting devices exhibit near-lossless characteristics, suppress thermal noise, and enable quantum coherence, making them appealing for next-generation computational hardware [5], [6]. Among these devices, superconducting nanowires (ScNW) have attracted significant interest over the past several decades [7]–[10]. Their ultrafast switching speeds, ultra-low power consumption, simple fabrication, CMOS compatibility, and natural suitability for cryogenic environments make them a compelling platform for compact and energy-efficient cryogenic electronic systems. To date, ScNWs have been utilized in a wide range of applications, including quantum computing hardware [11], space electronics [12], neuromorphic computing [13]–[19], etc.

Among the circuit components used in these applications, ScNW-based oscillators show promising characteristics due to their simplicity and unique oscillation dynamics [15]. These oscillators serve as critical building blocks for cryogenic systems, acting as timing sources, signal generators, or neuromorphic primitives.

Recent studies have demonstrated the viability of ScNW-based oscillator topologies for standalone applications, where voltage oscillations arise from nonlinear switching dynamics triggered by an external biasing scheme [15], [20]. Their minimalistic architecture enables scalable cryogenic signal generation, with potential use in systems such as neuromorphic processors and flux-controlled memory. However, to harness these oscillators for complex system-level tasks, such as neuromorphic computing or synchronized control in cryo-electronic systems, an in-depth understanding of their response to external stimuli and mutual coupling dynamics is essential.

Here, we extend our prior design space analysis of ScNW-based oscillators [15] by examining two notable dynamic behaviors: (1) the injection locking phenomenon driven by an external periodic current bias and (2) the coupling-induced synchronization dynamics between two ScNW oscillators. Injection locking—a phenomenon in which an internal oscillator synchronizes its frequency and phase with an external signal—has been widely studied in semiconductor and spintronic domains but remains to be comprehensively analyzed for ScNW-based oscillators. In this work, we examine the implications of critical device and circuit parameters on the overall injection locking dynamics. Our analysis identifies the critical design knobs for achieving tunability in the locking dynamics. It also generates necessary insights for low power, synchronized systems based on oscillators.

Furthermore, we present an in-depth investigation into the mutual coupling behavior of two ScNW-based oscillators connected via three different types of coupling elements: a capacitor, a resistor, and an inductor. We report the achievable phase difference between the coupled oscillators through the variation of the coupling strength of different coupling elements. This phase tunability introduces new design opportunities for implementing oscillatory neural networks (ONNs), phase-based logic circuits, and frequency-agile cryogenic resonators.

Through transient and frequency-domain simulations using our developed experimentally calibrated Verilog-A model, we expose the interplay between device and circuit parameters, such as shunt resistance, nanowire inductance, and coupling elements, and the resulting locking and coupling dynamics. These insights offer a pathway toward the co-design of materials, devices, and circuits for future cryogenic and neuromorphic computing systems.

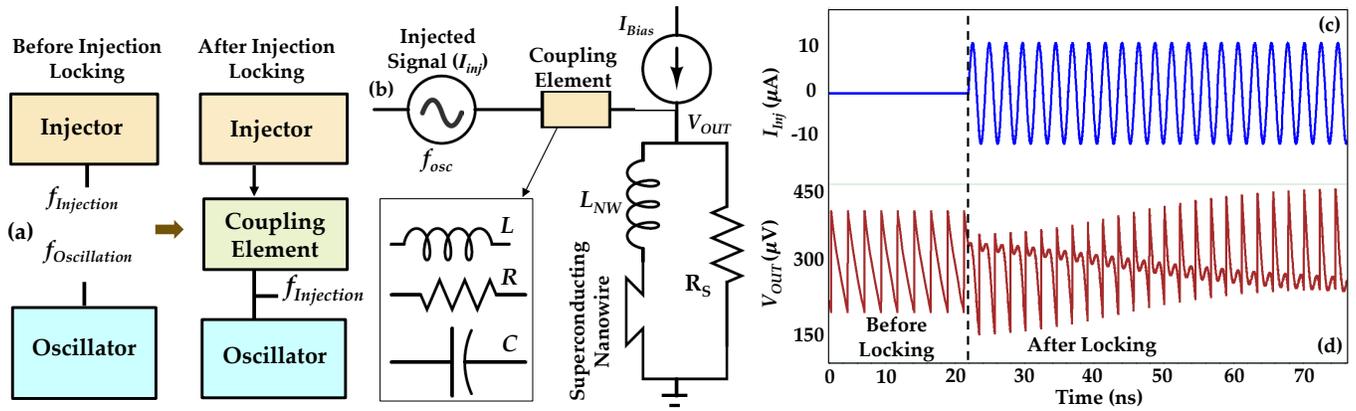

Fig. 1: (a) General mechanism of injection locking. An injection signal with frequency $f_{inj}$ is applied to a standalone oscillator operating at a frequency close to $f_{inj}$ through a coupling element. Upon successful locking, the oscillator frequency synchronizes to $f_{inj}$. (b) Injection locking in a superconducting nanowire-based cryogenic oscillator. Three types of coupling elements are considered in our analysis. (c) Injected sinusoidal current signal ($I_{inj}$), and (d) corresponding output voltage ($V_{OUT}$). Here, $f_{inj}$ = 443 MHz and $f_{inj}$ = 420 MHz. $V_{OUT}$ is coupled to $I_{inj}$ after successful synchronization.

The organization of this paper is as follows: Section II provides a brief introduction to the dynamics of the ScNW-based cryogenic oscillator structure. Section III discusses the injection locking dynamics and its comprehensive design space analysis. In Section IV, we discuss the coupling dynamics of ScNW-based cryogenic oscillators and comprehensively discuss the impact of coupling strength on the phase tunability of a pair of coupled ScNW-based oscillators.

## II. SUPERCONDUCTING NANOWIRE-BASED CRYOGENIC OSCILLATOR

We begin our discussion by delineating the structure and operating principles of the ScNW-based cryogenic oscillator. The core component of this design is a thin superconducting wire, referred to as a superconducting nanowire (ScNW). When a current is applied and increased gradually, up to a specific current threshold known as the critical current ($I_C$), the ScNW remains superconducting, maintaining zero voltage across it. However, when the current exceeds $I_C$, the ScNW transitions to a resistive state, causing a voltage to appear across the device. Once the current across the ScNW subsequently drops below a lower threshold, called the retrapping current ($I_r$), the ScNW re-enters its superconducting state. For the ScNW, $I_r$ is typically lower than $I_C$, resulting in hysteresis in their current-voltage (I-V) characteristics [15].

When the ScNW is shunted with a resistor ($R_S$), the resulting structure can produce an oscillatory voltage at the output. Upon current biasing, the entire current initially flows through the ScNW, and no current flows through the $R_S$, yielding a zero voltage at the output. However, when $I_{bias}$ exceeds $I_C$, the ScNW becomes resistive, causing the current to be redirected through $R_S$, producing a voltage rise at the output. This diversion reduces the current through the ScNW below $I_r$, allowing it to return to the superconducting state, which causes the voltage to fall again.

This repeating cycle generates sustained voltage oscillations, as reported in several earlier works. Here, the oscillation period is primarily governed by the nanowire inductance ($L_{NW}$) and the shunted resistance ($R_S$). The voltage rise time is determined by the time constant, $\tau_1 = L_{NW}/(R_S + R_{NW})$, where $R_S$ is the SM resistance, $R_{NW}$ and $L_{NW}$ represent the resistance and the inductance of ScNW, respectively. The fall time-constant is given by $\tau_2 = L_{NW}/R_S$. Thus, the oscillation period varies proportionally with the overall time constant ($\tau = \tau_1 + \tau_2$). A design space analysis of this oscillator, including the impact of material and device parameters, can be found in our earlier works [21].

## III. INJECTION LOCKING IN SCNW-BASED OSCILLATOR

Injection locking occurs when an external periodic signal (namely the "injector") forces the oscillator to synchronize to the injector's frequency once certain conditions are met (Fig. 1(a)). This occurs if the injector's frequency is sufficiently close to the oscillator's natural free-running frequency and the coupling strength exceeds a threshold value. This synchronization arises from nonlinear interactions within the oscillator's active components (e.g., transistors in LC-tank circuits), where the injector's energy suppresses the oscillator's autonomy.

To investigate the injection locking behavior of the ScNW-based oscillator, we introduce a sinusoidal current source as an external excitation (Fig.1(b)). Given the current-driven nature of the oscillator, a current-mode injection scheme is well-suited to probe synchronization dynamics. The injected signal has a frequency, $f_{inj}$, close to the natural (i.e., free running) frequency ($f_{osc}$) of the oscillator (Figs. 1(b,c)). This configuration allows us to explore classical injection locking behavior [22], where the oscillator can closely follow the phase and frequency of the external stimulus under appropriate conditions [23].

The injection current is coupled directly into the biasing path of the oscillator, forming a composite input that combines the DC bias current ($I_{bias}$) and the sinusoidal injection signal. The resulting current waveform forces the oscillator into a regime where its internal dynamics may align in both frequency and phase with the injected signal, depending on the amplitude and frequency detuning between $f_{inj}$ and $f_{osc}$.

Figs. 1(c,d) illustrate the time-domain waveforms of the injected signal and the oscillator output ($f_{out}$) before and after the locking. In the free-running condition, the oscillator exhibits a periodic voltage output determined by its intrinsic non-linear dynamics. Upon injection, the output waveform begins to

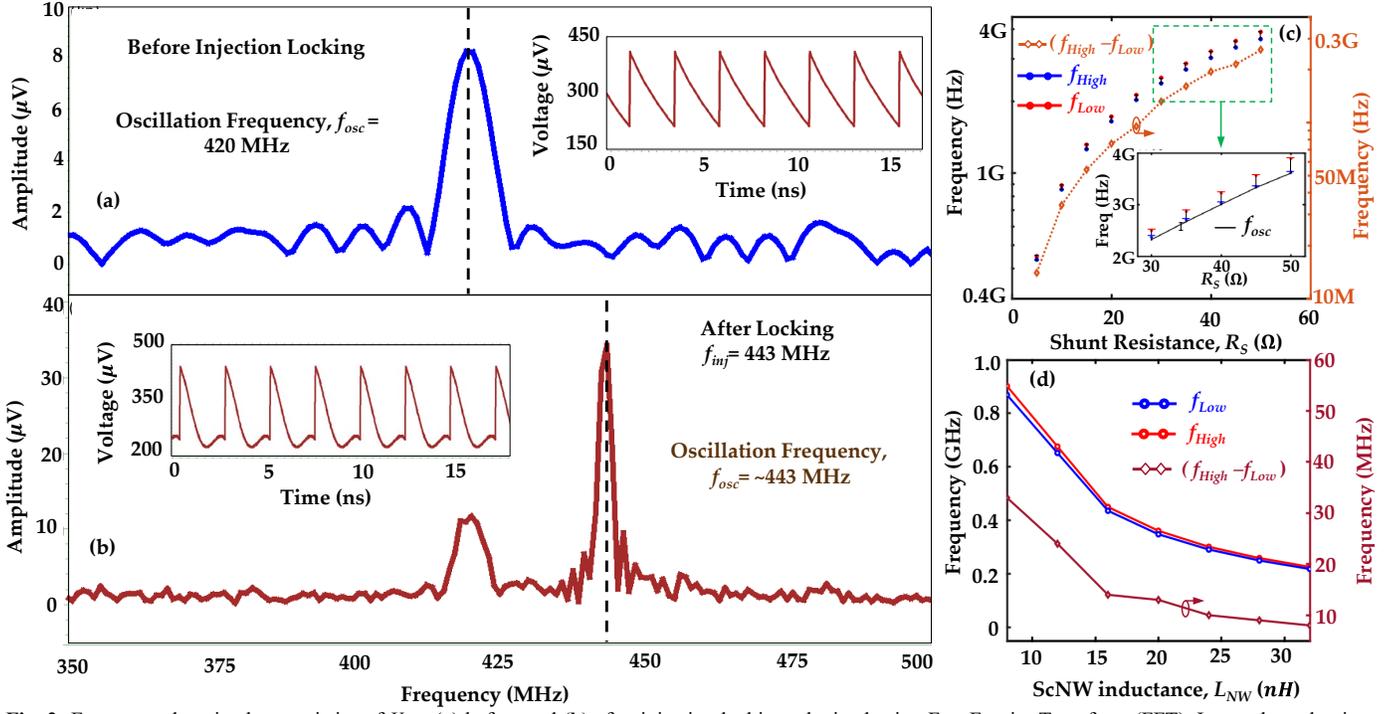

**Fig. 2:** Frequency domain characteristics of $V_{OUT}$ **(a)** before and **(b)** after injection locking, obtained using Fast Fourier Transform (FFT). Insets show the time domain characteristics for both before and after the locking. Here, the injected frequency is 443 MHz. The injected frequency, $f_{inj}$= 443 MHz. After successful locking, $V_{OUT}$ exhibits its dominant spectral peak at the injected frequency. **(c)** Variation of maximum locked frequency ($f_{High}$) and minimum locked frequency ($f_{Low}$) with shunt resistance ($R_S$). The locking range is defined as the difference between $f_{High}$ and $f_{Low}$. **(d)** Variation of $f_{High}$, $f_{Low}$, and the locking range with ScNW inductance ($L_N$).

exhibit phase alignment with the injected signnoal. When the injection frequency is sufficiently close to $f_{osc}$ and the amplitude is sufficiently large, the oscillator locks to the injected signal, as evidenced by a steady relationship between the two signals over time (Fig.1(d)).

To confirm the locking behavior, we perform frequency-domain analysis using 1024-point Fast Fourier Transform (FFT) [24]. Figs. 2(a,b) illustrate the spectral content of the oscillator output before and after the locking phenomenon. In the locked state (Fig. 2(b)), the oscillator's output spectrum exhibits a dominant peak very close to the injection frequency $f_{inj}$= ~443MHz, with a suppression of the amplitude of the inherent oscillation frequency, $f_{osc}$= 420 MHz during the unlocked state. The alignment of the oscillator's spectral peak with that of the injected signal serves as strong evidence of injection locking.

*A. Effect of Shunt Resistance and Nanowire Inductance:* We first examine the effect of shunt resistance ($R_S$) on the locking behavior. As shown in Fig. 2(c), increasing $R_S$ leads to a corresponding increase in the free-running oscillation frequency ($f_{osc}$) of the standalone oscillator. This trend is consistent with the relationship between $R_S$ and the overall oscillation period, as discussed in section II. For injection locking phenomenon, the locking range is defined as the range of frequency spectrum of the injected signal within which successful locking is achievable. As $f_{osc}$ increases, the center of the locking range also shifts higher. Notably, the locking range tends to increase slightly with higher $R_S$, indicating improved sensitivity to injection at higher oscillation frequencies. In contrast, increasing $L_{NW}$ lowers the natural frequency, $f_{OSC}$, leading to a downward shift of the locking range (Fig. 2(d)). A larger $L_{NW}$ lowers the natural oscillation frequency of the oscillator, $f_{osc}$ shifting the range of required injection frequency accordingly and the center of the locking range also decreases.

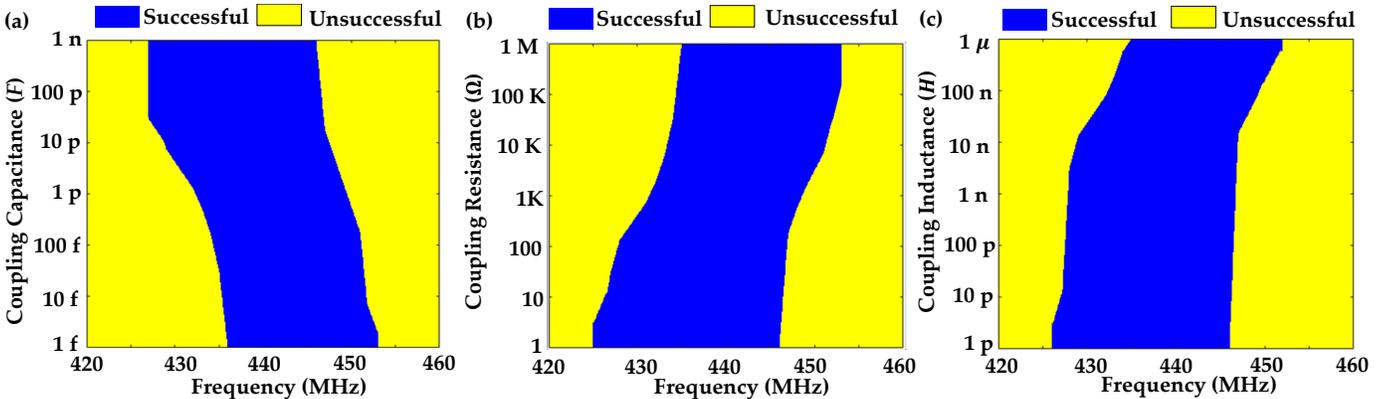

**Fig. 3:** Colormaps illustrating regions of successful and unsuccessful frequency locking for different coupling elements. Variation of locking range with **(a)** coupling capacitance, **(b)** coupling resistance, and **(c)** coupling inductance. In all cases, the locking range expands and shifts toward the natural oscillation frequency ($f_{inj}$) with the increase of the coupling strength.

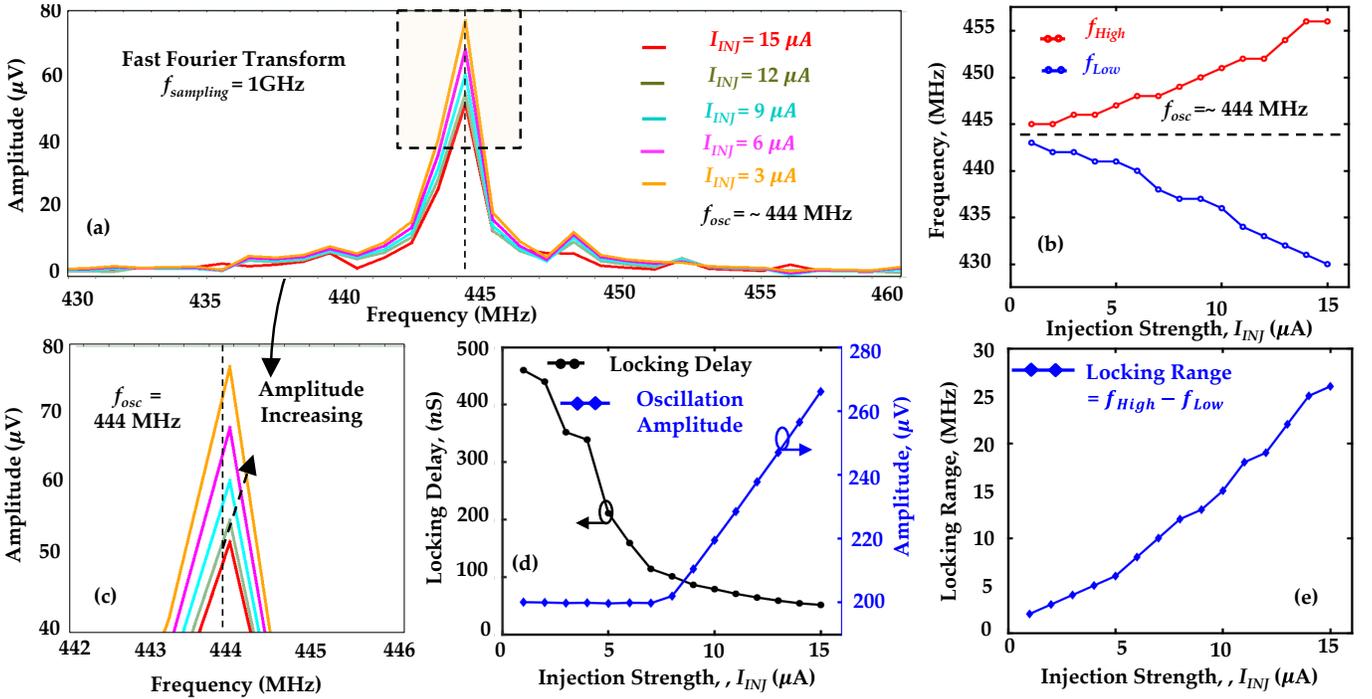

**Fig. 4:** Impact of injection amplitude on injection locking in the ScNW-based oscillator. **(a)** Frequency-domain characteristics of the locked oscillator output for five different injection strengths: 3μA, 6μA, 9μA, 12μA, 15μA. **(b)** Variation of maximum locked frequency ($f_{High}$) and minimum locked frequency ($f_{Low}$) with the change of injection strength/amplitude. **(c)** Zoomed-in view of the rectangular region highlighted in panel (a). **(d)** Dependence of locking delay and locked oscillation amplitude on injection strength. **(e)** Locking range vs injection amplitude, showing a linear trend ($\Delta f \propto I_{inj}$).

This analysis establishes the fundamental behavior of the ScNW oscillator under periodic perturbation and lays the groundwork for examining how key circuit parameters—such as shunt resistance, nanowire inductance, and injection strength—influence the optimal locking range. These dependencies are explored in detail in the following section. Notably, for very large values of $L_{NW}$, we observe a narrowing of the locking range, indicating that increased inductance leads to slower oscillatory dynamics, which in turn hinder the oscillator's ability to respond effectively to external perturbations. Outside the locking range, when the frequency detuning $|f_{inj} - f_{OSC}|$ exceeds a critical threshold, the oscillator either exhibits amplitude modulation due to partial synchronization or fails to respond altogether.

*B. Effect of Coupling Strength on the Locking Range:* Next, we investigate the impact of the coupling element on the locking range. As shown in Figs. 3(a-c), we explore three types of coupling elements—resistors, capacitors, and inductors—and sweep their values to evaluate how the locking range evolves. In each case, we observe a consistent trend: increasing the coupling strength—via higher capacitance, lower resistance, or lower inductance—broadens the frequency range over which the oscillator can lock to the injected signal frequency. This behavior is consistent with classical locking theory [22], which predicts that the locking range scales with coupling strength. Stronger coupling facilitates more efficient energy transfer from the injector to the oscillator, allowing the overall coupling range to shift more towards the injection frequency ($f_{inj}$). This is achieved by altering the oscillator's input impedance, enabling greater modulation of its internal current waveform by the external injection source.

*C. Effect of Injection Strength on Coupling Range and Coupling Delay:* Next, we investigate the impact of injection strength on the overall injection locking dynamics (Fig. 4). As the injection strength (or amplitude) increases, the amplitude of the oscillator's output in the locked state also increases, as shown in Figs. 4(a,d). In the frequency-domain spectra illustrated in Figs. 4(a,c), the spectral peak becomes increasingly pronounced with the increase of the injection frequency, while changes in amplitude at other harmonic frequencies remain minimal. For example, the amplitude of the second harmonic (888 MHz) increases by less than 1% when the injection strength is increased by 12 μA (from 3μA to 12 μA).. Fig.4(b) illustrates the upper ($f_{High}$) and lower ($f_{Low}$) bounds of the injection frequency range over which successful locking is achievable.

As shown in Fig. 4(b), the injection amplitude directly impacts both $f_{High}$ and $f_{Low}$. Consequently, the overall locking range linearly increases with the injection amplitude (Fig. 4(e)). This trend agrees with the classical injection locking theory, where the locking range linearly varies with the injection strength ($\Delta f \propto I_{inj}$) [25]. Moreover, we examine how the injection amplitude impacts the locking delay-the characteristic time required for the oscillator to synchronize with the injected signal. As depicted in Fig. 4(d), higher injection strengths lead to shorter locking delays, indicating fa aster convergence to the locked state under stronger coupling conditions. For instance, increasing the injection amplitude from 1 μA to 15 μA results in more than a 90% reduction in locking delay.

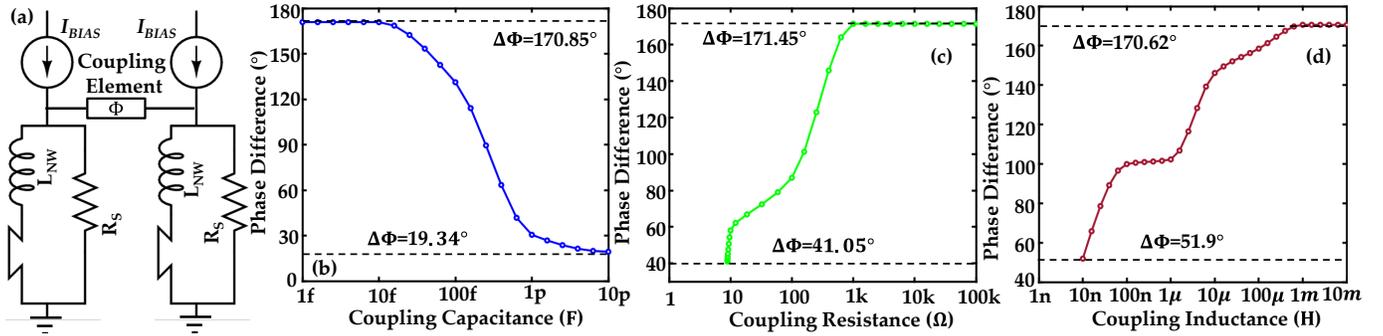

Fig. 5: (a) Schematic of two ScNW-based oscillators coupled via a coupling element to achieve a tunable phase difference (ΔΦ) between their oscillations. Variation of the phase difference (ΔΦ) with varying coupling strength for (b) capacitive coupling, (c) resistive coupling, and (d) inductive coupling. The dashed lines mark the maximum and minimum achievable ΔΦ for the corresponding coupling elements.

Collectively, these results underscore the critical role of circuit and material-dependent parameter ($L_{NW}$) selection in enabling effective injection locking in ScNW-based oscillators. The ability to tune the locking range through $R_S$, $L_{NW}$, injection strengths, and coupling elements provides a set of physical design knobs for achieving robust and frequency-locked operation in cryogenic systems. These findings are particularly relevant for applications where oscillator synchronization must be dynamically controlled, such as in oscillatory neural networks for adaptable signal processing or where specific locking ranges are desirable for system robustness, as in cryogenic clock distribution networks [26]–[29].

## IV. COUPLING MECHANISM IN ScNW-BASED OSCILLATOR

Next, we investigate the coupling dynamics between a pair of ScNW-based oscillators. As illustrated in Fig. 5(a), the two oscillators are interconnected through a coupling element. Identical device and circuit parameters, along with equal bias currents ($I_{bias}$), are used to ensure both oscillators operate at the same natural frequency ($f_{osc}$). Our goal is to achieve a tunable phase difference (ΔΦ) between the oscillators by varying the coupling strength across different types of coupling elements—capacitor, resistor, and inductor. The objective is to achieve a tunable phase difference (ΔΦ) between the oscillators by varying the coupling strength across different types of coupling elements—capacitive, resistive, and inductive. Since the oscillators generate identical waveforms at the same frequency, the phase difference is quantified by measuring the time offset between their nanowire switching events, normalized to the oscillation period. Fig. 5(b) shows the evolution of ΔΦ as a function of coupling capacitance. At low capacitance values (weak coupling), the oscillators are nearly anti-phase, exhibiting a high phase difference. As the capacitance increases (stronger coupling), the phase difference decreases progressively. A limit to the tunable phase range is observed (ΔΦ=170.85°, ΔΦ=19.34°) and indicated by the dashed lines in Fig. 5(b). In the strong coupling regime, the two oscillators begin to act as a single synchronized system, indicating a transition from independent to collective behavior.

A similar phase-tuning behavior is observed for resistive coupling, as shown in Fig. 5(c). At high resistance (weak coupling), the phase difference reaches up to ~171.45°. As resistance decreases and coupling strengthens, the phase difference reduces, reaching a minimum of ΔΦ=41.05°, illustrating the transition from weak to strong coupling. An analogous trend is seen with inductive coupling in Fig. 5(d). Lower inductance values yield stronger coupling and smaller phase differences, while increasing inductance weakens the coupling, allowing ΔΦ to rise up to 170.62°.

Across all three coupling mechanisms—capacitive, resistive, and inductive—we observe a strong coupling regime in which the two oscillators behave as a single, unified system. In this regime, energy is rapidly and periodically exchanged, leading to the formation of collective oscillation modes (normal modes) with periods distinct from the uncoupled systems. These results demonstrate that the phase difference between ScNW-based oscillators can be precisely controlled over a wide range by tuning the coupling element's type and strength. This tunability holds strong potential for applications in synchronized oscillator systems, combinatorial optimization solvers, and oscillatory neural networks (ONNs) [28], [30], where information can be encoded in the phase difference between coupled oscillators.

## V. CONCLUSION

Here, we have conducted a comprehensive simulation-based study on the injection locking and mutual coupling dynamics of superconducting nanowire (ScNW)-based cryogenic oscillators. Our findings offer new insights into the behavior of these oscillators under external periodic excitation and mutual interaction-two critical features for the realization of scalable cryogenic computing systems and neuromorphic architectures[31]. Through detailed transient and frequency-domain analyses, we demonstrate that the injection locking behavior in ScNW oscillators is strongly influenced by several circuit parameters. While our results are consistent with classical injection locking theory, they represent the first such validation in the context of ScNW-based cryogenic devices.

We also investigated the impact of injection strength (amplitude) on the locking dynamics. This analysis has important implications for energy-aware designs, where robust synchronization must be achieved with minimal injected power.

Beyond single oscillator locking, we have explored the mutual coupling between two ScNW oscillators. By introducing capacitive, resistive, and inductive coupling elements, we demonstrated tunable phase differences through modulation of coupling strength. Such precise phase control enables the implementation of phase-based information processing schemes, including phase-encoded logic circuits and oscillatory neural networks (ONNs) [28], [30], [32], where computational states are encoded in phase relationships. Our analysis reveals that above a critical coupling threshold, the oscillators behave collectively, exhibiting new oscillation modes distinct from their uncoupled natural frequencies.

Overall, this study provides a foundational understanding of synchronization dynamics in ScNW-based oscillators and identifies key design knobs for controlling their collective behavior. These findings will guide the development of future cryogenic computing architectures that leverage the inherent energy efficiency and high-speed dynamics of superconducting nanowires. Future work should focus on experimental validation and scaling toward large oscillator networks for solving complex optimization problems and enabling real-time neuromorphic computing at deep cryogenic temperatures.

## Acknowledgement

The work was supported by the Science Alliance, a Tennessee Higher Education Commission Center of excellence administered by the University of Tennessee-Oak Ridge Institute on behalf of the University of Tennessee Knoxville. This work was supported in part by the U.S. Department of Energy, Office of Science, Advanced Scientific Computing Research (ASCR) program, under Award DE-SC0024328.